\documentclass{article}
\usepackage{graphicx}
\usepackage[latin1]{inputenc}
\usepackage{amsmath}
\graphicspath{{./}{./figures/}}

\advance\textwidth by 3cm
\advance\oddsidemargin by -1.5cm
\advance\evensidemargin by -1.5cm

\begin{document}
\title{\bf Pushing the complexity barrier: \\ diminishing
        returns in the sciences}
\author{Claudius Gros \\
  Institute for Theoretical Physics, \\
  J.W.\ Goethe University Frankfurt, Germany} 
\maketitle
Are the sciences not advancing at an ever increasing speed?
We contrast this popular perspective with the view that
scientific research is actually closing in to complexity barriers
and that, as a consequence, science funding actually sees 
diminishing returns, at least in established fields. In order to       
stimulate a larger discussion, we investigate two exemplary cases,
the linear increase in human life expectancy over the last
170 years and the advances in the reliability of numerical
short and medium term weather predictions during the last
50 years. We argue that the outcome of science and technology
funding in terms of measurable results is a highly 
sub-linear function of the amount of resources committed.
Supporting a range of small to medium size research projects,
instead of a few large ones, will be, as a corollary,
a more efficient use of resources for science funding agencies.

\subsection*{Measuring scientific progress}

There is a curious dichotomy in our current science 
and technology (S\&T) landscape. On one side we see 
advances on scales as never before in human history. 
There is, on the other side, a growing sentiment among 
researchers that progress in science is becoming 
increasingly harder to achieve. This sentiment is based 
in part on anecdotal evidences and is continuously reinforced
by new insights. On the anecdotal side there is the
phenomenon that the requirements for a typical PhD thesis 
in the natural sciences have increased dramatically over 
the last 50 years. It is well known that it takes nowadays
much longer for a young scientist to reach the forefront
of research. 

The notion that scientific research needs to deal with
raising levels of complexity is especially evident when 
studying the realm of life. A paradigmal example of new 
insights bolstering this notion are the results of the 
ENCODE project \cite{encode12}, showing that our genome 
does not only contain 21 thousand protein encoding genes, 
but also up to 4 million regulatory switches where transcription 
factors could bind, besides a myriad of other regulatory
sequences. Is it possible to quantify this notion of
a raising complexity level? This is the central topic
of our investigation and it involves the quest to
actually measure the pace of scientific progress.

Scientific progress is notoriously hard to measure.
It doesn't really make sense to quantify advances in 
fundamental research; a single publication leading to 
a paradigm shift may be invaluable. The vast majority 
of scientific research efforts are however directed 
towards achieving incremental progress, and are not of 
foundational character. It is hence worthwhile to ask how 
taxpayers' money allocated for public science funding 
could be spent most efficiently.

In order to make a first inroad we investigate two large-scale
endeavors of humanity. The first case study concerns the 
long-term impact of research and investments, in medicine and 
healthcare, on life expectancy over the last 170 years, asking
the somewhat antipolar question: Why did the average life
expectancy rise so slowly? The second example regards the
advances in the reliability of short and medium term
numerical weather forecasts since the 1950s. Weather dynamics
has potentially chaotic regimes and the pace of progress
in predictive meteorology may be limited by a complexity 
barrier resulting from systemic difficulties in predicting 
chaotic dynamical systems.

We find that measurable progress in S\&T is a highly 
sub-linear function of the invested resources, reflecting
the law of diminishing returns well studied in economical
contexts \cite{marshall09}. Some scientific insights
can be achieved only through large collaborative projects,
like the search for the Higgs Boson. Our results however
show that small scientific endeavors do generically offer
a higher potential for returns in terms of results per 
allocated funding.

Many natural systems investigated in the sciences are complex
dynamical systems \cite{gros08}. The brain in the neurosciences
and the human genome in bioinformatics are examples from the
realm of life. Short- and long term weather and climate 
evolution, many-body systems in condensed matter physics 
and elementary particle condensates are examples from the 
realm of physics. Complex systems are both difficult to
understand and to investigate on a conceptual basis as well
as to model and simulate numerically. These two difficulties
impinge the pace of progress when investigating complex
biological, or physical systems. We find the
notion of a malleable complexity barrier to be a good
visualization for the challenges confronting today's 
scientists.

\begin{figure}[t]
\centerline{
\includegraphics[width=0.70\textwidth]{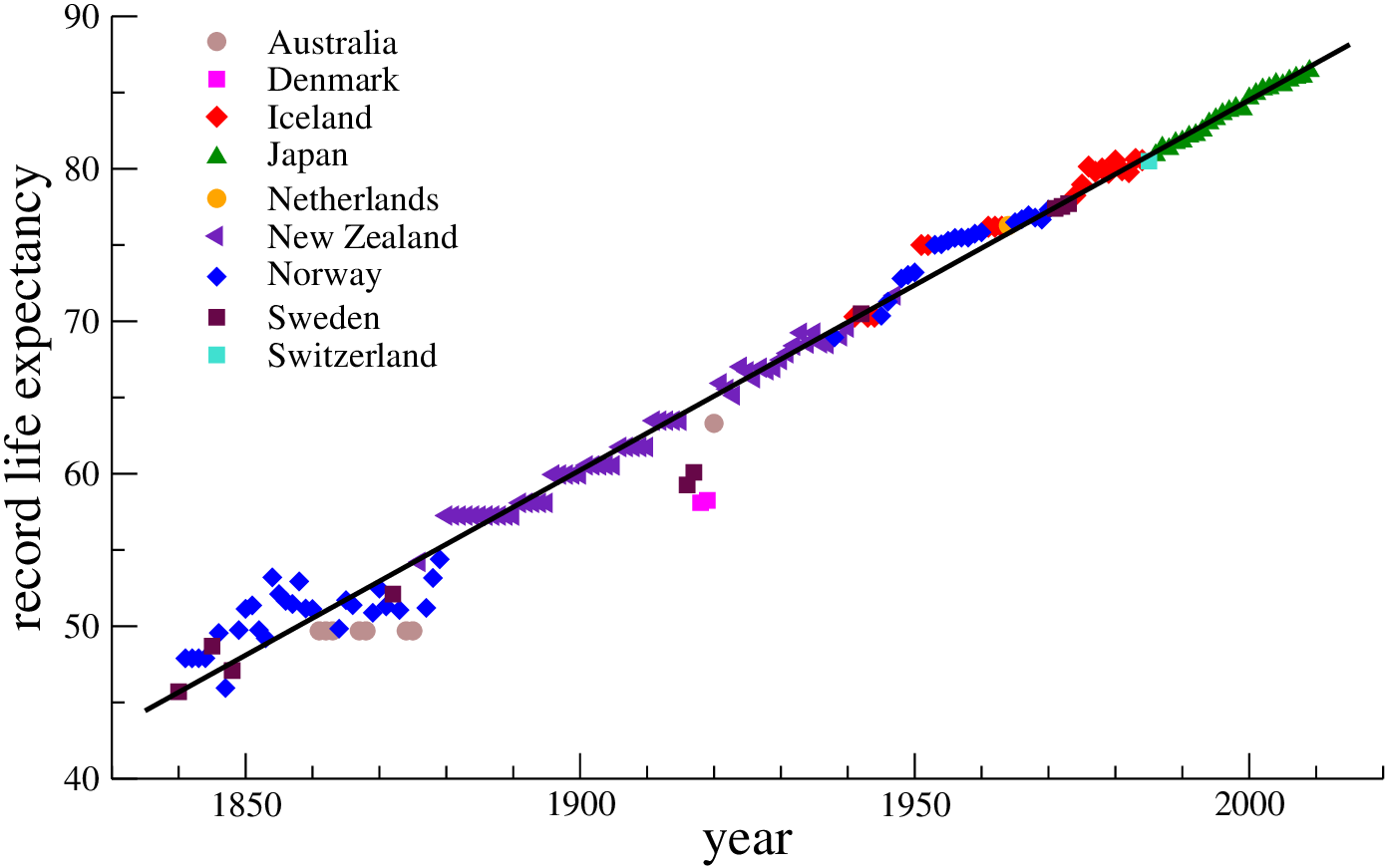}
           }
\caption{Record female life expectancy \cite{oeppen02,human}. 
         Shown is the life expectancy of the country with the 
         highest average female life expectancy in a given year, 
         for all calendar years.
         The line is a least square linear regression,
         the increase in record life expectancy averaging
         2.4 years for 10 calendar years.
        }
\label{fig_life}
\end{figure}

\subsection*{Record life expectancy}

Human life expectancy has seen a dramatic rise since
the mid of the 18th century. On a global level one
commonly considers the 'record life expectancy' as the
life expectancy at birth of the country having the highest
life expectancy worldwide. The record life expectancy has 
seen a strikingly linear growth for about 1.5 centuries, as 
shown in Fig.~\ref{fig_life}. It has been repeatedly predicted
that this steady increase of human life expectancy would
need to level off at a certain point, invoking biological 
limits. All these predicted limits have been broken hitherto 
without exception \cite{oeppen02,white02}. This spectacular 
steady growth of the record life expectancy raises a series 
of interesting points.

Advances in healthcare and medicine will lead generically to
raising levels of life expectancy. It remains however 
unclear which forces determine the magnitude of the observed 
rate, 2.4 years per every 10 calendar years, and how the
observed growth rate depends on the overall 
amount of resources devoted \cite{tuljapurkar00}. 
Breakthroughs in research have been postulated on
one hand to boost human lifespans rapidly \cite{deGrey03},
a putative natural limit, if existing \cite{oeppen02,vijg05},
should lead on the other side to a gradual levelling off.
We have yet no definite answers to these foundational 
questions.

Longevity is a central issue in our culture and major
efforts and resources are devoted by our societies
towards increasing health levels and lifespan.
Fig.~\ref{fig_life} demonstrates that returns on
investments have dramatically decreased during the last
1.5 centuries. The initial growth in life expectancy
resulted from simple hygiene measures, followed by
progress in immunology and antibiotics research.
Lately, massive investments in pharmacology, technical
medicine and bioinformatics have been necessary to
keep up the steady linear advance in life expectancy.
Relative progress has actually decreased in spite of 
these massive efforts, a linear increase relative to a base 
of 45 years is twice as large as a linear increase 
(with the same slope) relative to a base of 90 years.

\begin{figure*}[t]
\centerline{
\includegraphics[width=0.70\textwidth]{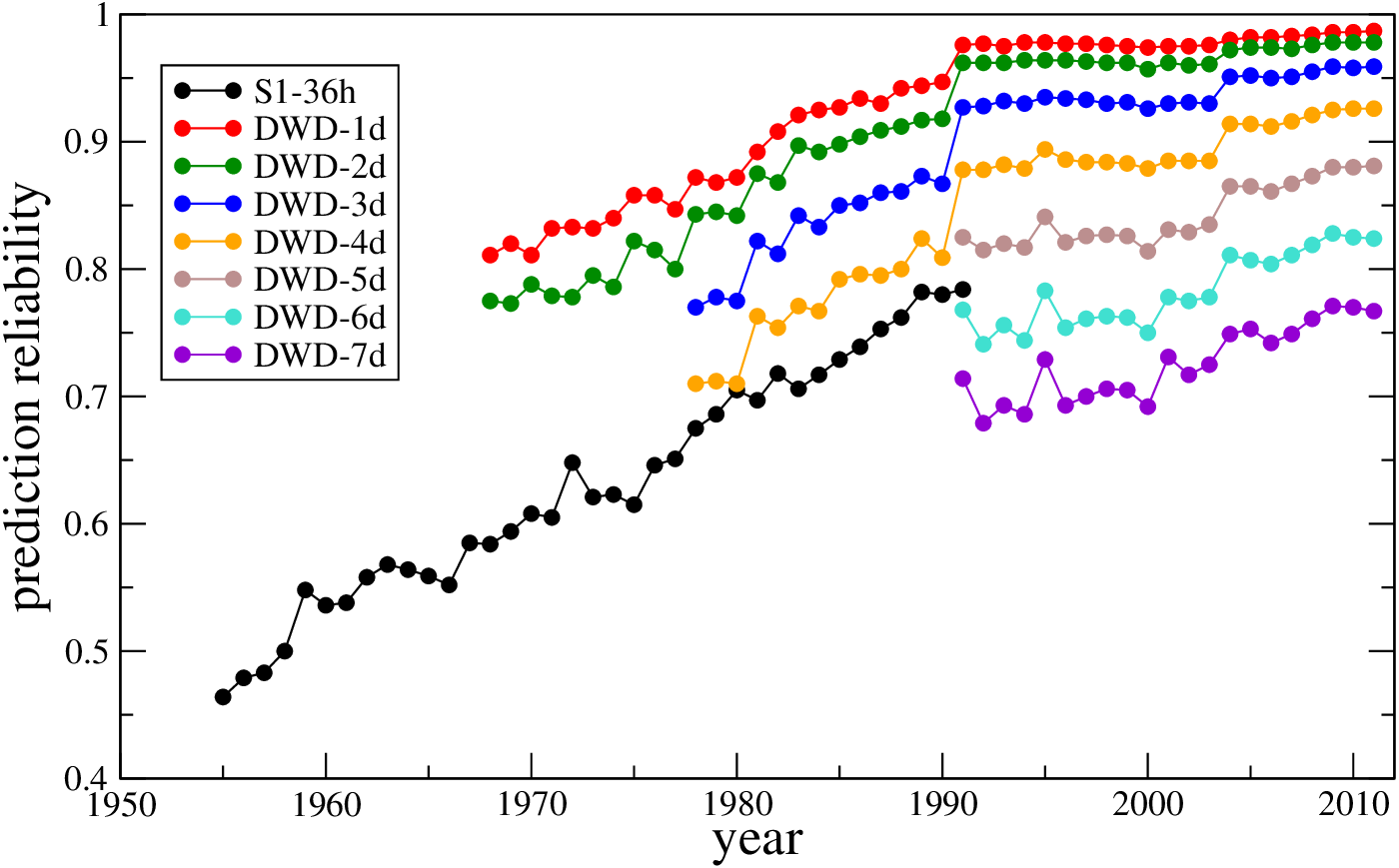}
           }
\caption{Shown are two distinct weather forecast reliability 
         measures, the 500MB 36h (hour) S1 score of the NOAA
         \cite{NOAA92} (black data), and the 1d-7d (days) 
         500hPa correlation coefficient of the DWD \cite{DWD12}
         (color data). Note that these two reliability scores 
         are differently defined and cannot be compared 
         directly on a quantitative basis.
        }
\label{fig_meteo_bare}
\end{figure*}

Investment in medicine and healthcare have seen
seen a roughly exponential increase since at least 
half a century \cite{anderson00,sisko09}. The
driving forces behind these ever raising costs for
healthcare and medicine are debated and could be
rooted either in the desire to increase health and 
well-being quite generally or, more directly, in the quest 
to postpone death as far as possible. It has been 
argued, in this context, that the economic rational 
behind the ever raising levels of health spending 
lies in the fact that humans attribute an income 
elasticity well above unity to improvements of 
life expectancy, which seems to have psychologically
a non-declining marginal utility \cite{hall04}.
This argument indicates that the average life-expectancy 
is indeed a valid yardstick for measuring the
overall advancement in health and medicine.
The quantitative efficiency of S\&T research efforts 
in the healthcare sector is hence, as measured by the 
observed extension of human life expectancy, a highly 
sub-linear function of resources devoted.
Returns on investment are seen to diminish rapidly in
medicine and healthcare.

\subsection*{Weather forecast reliability}

The dynamics of weather and climate for medium to long
time scales is known to contain chaotic components 
\cite{shukla98}. Indeed, one of the central models in the 
theory of chaotic and complex systems, the Lorenz model 
\cite{lorenz63}, is a hydrodynamic convection model.
Medium to long term weather forecast is hence considered
a challenge and massive investments in modeling,
data acquisition and computational infrastructure have
been made in order to achieve improved forecast performances.

A range of forecasting skill scores for numerical
short to medium term weather prediction are evaluated 
continuously, in order to track the quality of daily 
weather predictions, by national and cross-national 
weather and climate agencies. In 
Fig.~\ref{fig_meteo_bare} the historical evolution
of two distinct prediction reliability measures are 
shown, the first is the 1-7 day 500hPa forecast 
correlation coefficients of the DWD 
(Germany Weather Service) \cite{DWD12,murphy89}.
A value close to unity signals optimal forecasting, 
values below 60\% corresponding to essentially useless 
predictions. The introduction of a new
model in 1990 is reflected in the data. Also given in
Fig.~\ref{fig_meteo_bare}, for a longer-range
perspective (1955-1991), is the S1 36 hour skill score 
of the NOAA (National Oceanic and Atmospheric
Administration) \cite{NOAA92,lynch08}, which 
we normalized to the interval [0,1].
The S1 skill score is qualitatively different,
containing gradients, from the 500hPa correlation 
coefficient \cite{lynch08}, a direct comparison of 
the absolute values is not meaningful for these to
weather forecasting reliability measures.

\begin{figure*}[t]
\centerline{
\includegraphics[width=0.70\textwidth]{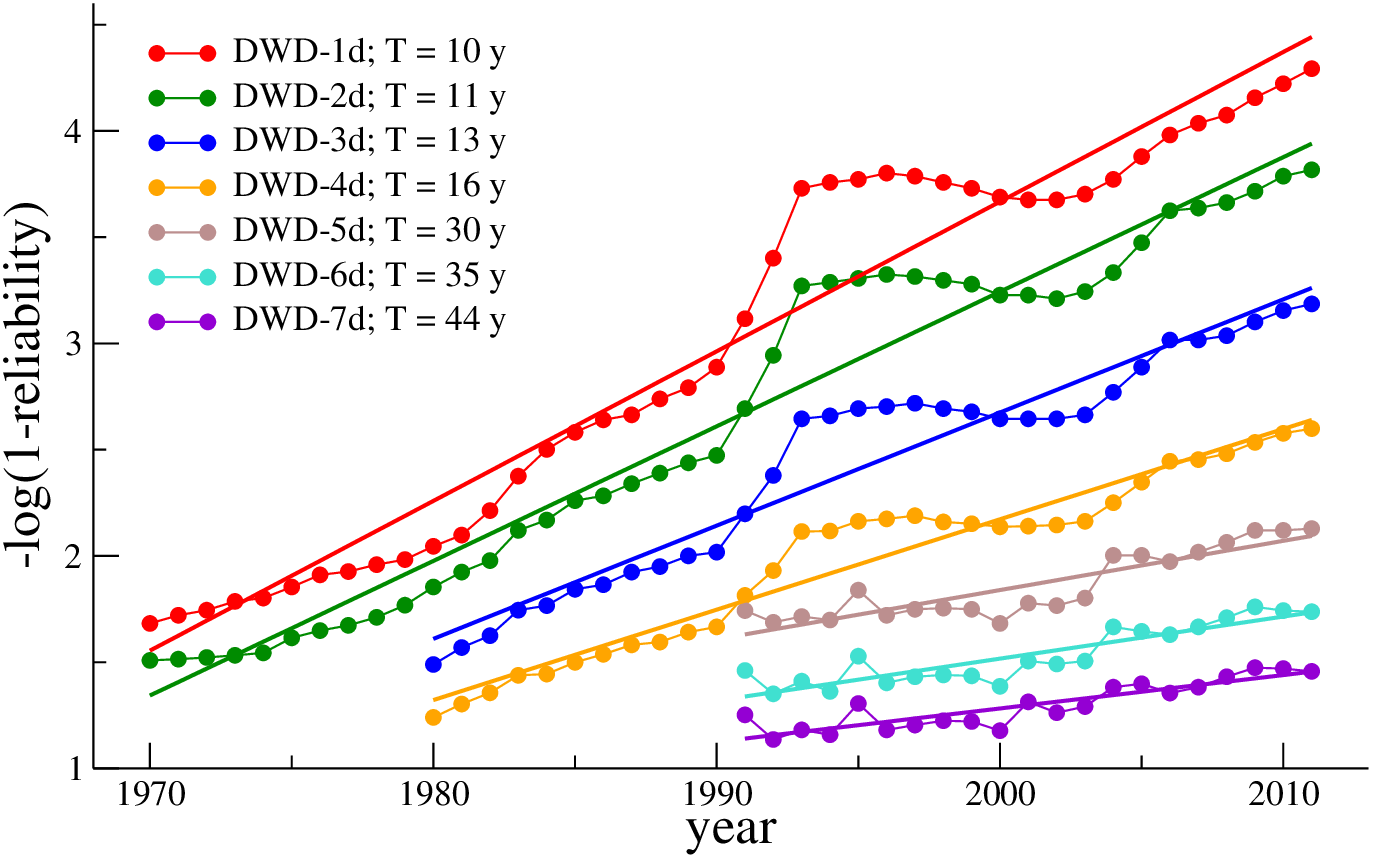}
           }
\caption{Linear-log plot of the DWD data (three years
         trailing average) of Fig.~\ref{fig_meteo_bare}. 
         The lines are least square 
         regressions, the inverse slope in terms of the number
         $T$ of years needed to double the relative precision 
         is given in the legend.
        }
\label{fig_meteo_log}
\end{figure*}

One of the key ingredients for numerical weather prediction, 
besides modeling and data acquisition \cite{lynch08},
is computing power. Available computing power has seen 
an exponential growth over the last 60 years, Moore's law 
\cite{moore98,thompson06}, with a doubling period of about 
1.5 years. The resulting advances in computational power 
has been about $10^{10}$ in 50 years. The computational 
facilities employed for numerical weather forecasting 
have seen corresponding increases \cite{lynch08},
contributing to the observed improvements in weather
forecasting skills.

In order to estimate the scaling between computational
resources and forecasting skills quantitatively, we show 
in Fig.~\ref{fig_meteo_log} the standardized forecasting 
error, $(1-\mbox{reliability})$, corresponding to the
remaining difference to optimal forecasting.
The S1 skill score  data is too distant from optimality,
we hence focus on the DWD data for a systematic analysis.
In Fig.~\ref{fig_meteo_log} we have reploted the 
DWD data as $-\log(1-\mbox{reliability})$, using 
a trailing 3-year average as a smoothing procedure.
In order to compare the growth rates of prediction accuracy
for different forecasting timescales we analyze the
data presented in Fig.~\ref{fig_meteo_log} using least
square regressions. They fit the data reasonably well,
indicating that the long-term growth of prediction 
accuracy is roughly exponential.

\begin{figure*}[t]
\centerline{
\includegraphics[width=0.70\textwidth,angle=0]{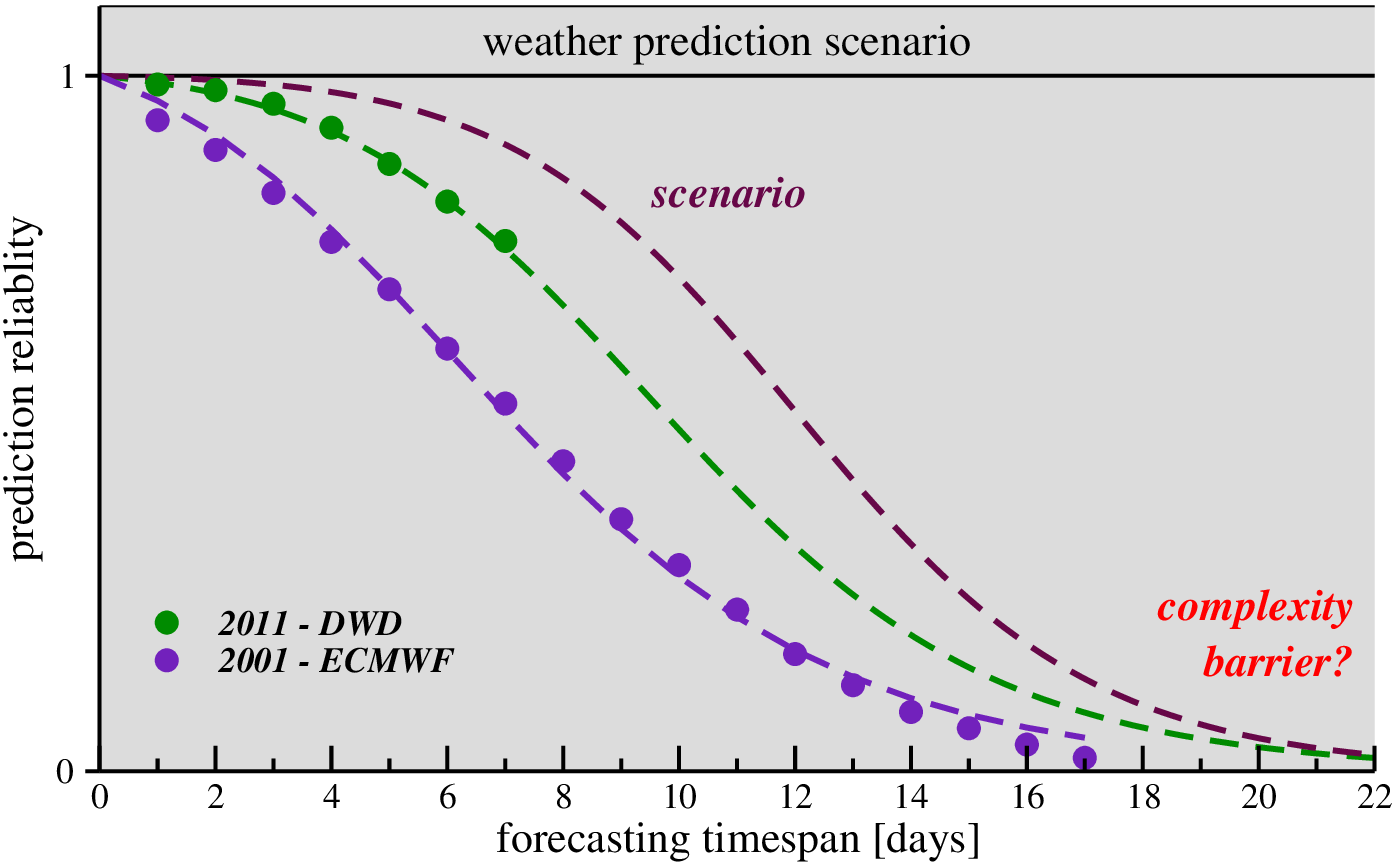}
           }
\caption{Forecasting reliability of the ECMWF (European Center 
         for Medium-Range Weather Forecasts) T255L40 score from
         2001 \cite{simmons02} (violet data), renormalized to
         $[0,1]$, and of the 2011 DWD 500hPa score (see 
         Fig.~\ref{fig_meteo_bare}, green data). 
         The dashed lines are two-parameter fits using
         Eq.~(\ref{eq_fit}), intended as guides to the eye. 
         The brown dashed line represents a putative scenario 
         for reliability scores achievable with further advancements.
        }
\label{fig_meteo_scenario}
\end{figure*}

The time needed to double the relative accuracy, i.e.\ 
to half the the forecasting error $(1-\mbox{reliability})$,
grows systematically with increasing forecasting timespan 
(see the legend of Fig.~\ref{fig_meteo_log}). The
accuracy of one-day weather predictions has doubled
historically roughly every 10 years. Of the order of
about 40 years seem to be necessary, on the other hand,
for improving the reliability of seven day weather 
forecasting by a corresponding factor. Out of these
results we conclude the following:

Firstly, the quantitative progress in weather prediction
accuracy is a highly sub-linear function of committed
computing resources. The accuracy scales roughly, with 
respect to the power $P_c$ of the computer facilities 
employed, as
$\left(P_c^{1.5}\right)^{1/10}= P_c^{0.15}$ 
for one day forecasting and as
$\left(P_c^{1.5}\right)^{1/40}= P_c^{0.0375}$ 
for seven day predictions. The reliability of the skill 
scores is dependent additionally on advances in modeling
and data acquisition, the respective functional relations 
of these dependencies is however beyond the scope of the 
present discussion.

Secondly, forecasting becomes rapidly more difficult
with increasing prediction periods. Indeed it
has been suggested that it may essentially be
impossible to achieve useful forecasting reliabilities
for 14-21 days in advance \cite{simmons02}, at least 
with economically justifiable amounts of resources.
This can be seen by plotting the weather forecasting scores 
as a function of forecasting period, as done in 
Fig.~\ref{fig_meteo_scenario}, where we have included
also the T255L40 reliability score of the ECMWF 
(European Center for Medium-Range Weather Forecasts) 
from 2001 \cite{simmons02}. Also shown in
Fig.~\ref{fig_meteo_scenario} are guides to the 
eye in the form of two-parameter least-square fits
of the functional form 
\begin{equation}
1-\frac{t}{t+a\exp(-bt)}~,
\label{eq_fit}
\end{equation}
for $t=1,\,2,\,\dots$ days, with adjustable parameters 
$a$ and $b$. This functional form captures the notion 
that it becomes progressively more difficult to achieve 
reliable forecasting with increasing prediction periods, 
the functional form of this increase in complexity has been 
assumed here to be exponential.

Generalizing these two observations we may relate
the measured S\&T progress to the amount of committed
resources via the scaling relation
\begin{equation}
\mbox{progress} \ \propto\ (\mbox{resources})^\alpha~,
\label{eq_scaling}
\end{equation}
with a sub-linear scaling exponent $\alpha<1$.
Our results, Figs.~\ref{fig_meteo_log}
and \ref{fig_meteo_scenario}, indicates,
that the scaling exponent $\alpha$ rapidly drops 
towards zero with increasing complexity of the task 
to be solved. We propose
to use the term {\em `complexity barrier'}
for this phenomenon. For the
case of the extension in human life expectancy,
which is raising linearly, invested
resources have increased roughly exponentially
\cite{anderson00,sisko09}, leading to a logarithmic
relation, $\mbox{progress} \propto \log(\mbox{resources})$.
A log-relation corresponds to a vanishing scaling
exponent, $\alpha\to0$, indicating that increasing the
average human life expectancy is a task of very high
complexity. 

\subsection*{Entering the human factor}

The complexity barrier present in the context of short 
to medium term weather forecast is not hard. Progress 
is achievable when committing increasingly larger amounts 
of resources. The same holds for the complexity barrier 
present in the aging problem. Extending the average human 
life expectancy has been possible for the last 170 years
by devoting strongly increasing amounts of resources to 
medicine and healthcare. The magnitude of the resources 
committed has been growing not only in absolute terms, 
but also as a fraction of national GNPs. The rational for 
the underlying collective decision of resource distribution is to be 
attributed to the human factor, progress in extending the
human life span is highly valued.

We have discussed here only two examples, but we believe
that progress in science, whenever it can be measured
on a quantitative basis, is quite generically a strongly 
sublinear function $f(x)$ of the amount $x$ of 
resources devoted. Sublinear dependencies are concave 
and for any concave function $f(x)$
the total return $\sum_if(x_i)$ is larger when splitting 
the total amount of resources $x$ into a series of
subpackages of smaller sizes $x_i$,
$$
f(x) \ <\ \sum_if(x_i), \qquad \sum_i x_i =x~.
$$
It constitutes hence, for funding agencies, a substantially
more efficient use of available resources to prioritize 
small to medium size projects, whenever feasible.

It is interesting to speculate, on a last note, whether
the human factor influences the pace of progress in
S\&T in an even more direct way. The human brain is well
known to discount incoming information streams 
logarithmically, a relation known as the Weber-Fechner law
\cite{hecht24,nieder03,dehaene03}. This exponential data
compression is necessary in order not to drown in the
daily flood of sensory impressions. It has been 
observed recently, that these neuropsychological constraints 
shape the statistics of global human data production in
terms of publicly available data files in the 
Internet \cite{gros11}. The human factor is hence
in evidence, at least in this particular aspect of human 
S\&T efforts, in the statistics of global public data generation.
It is hence conceivable, as a matter of principle, 
that the pace of progress close to a complexity barrier 
is not only influenced by the overall amount of financial 
and physical resources committed, but also more directly 
by the neuropsychology of human thought processes.

I would like to thank Ulrich Achatz, George Craig, 
Ulrich Damrath, Peter Lynch, Detlev Majewski, Peter Nevir 
and Johannes Quaas for support collecting the 
meteorological skill scores, Roland Rau and James Vaupel for the
life expectancy data. 


\end{document}